\documentclass{emulateapj}
\usepackage{epsfig}
\usepackage{natbib}

\def\deg{\ifmmode {^\circ}\else {$^\circ$}\fi} 
\def\arcmin{\ifmmode^{\prime~}\else $^{\prime~}$\fi} 
\def\arcsec{\ifmmode^{\prime \prime~}\else $^{\prime \prime~}$\fi} 
\def\msun{\ifmmode {\rm M_{\odot}}\else $\rm M_{\odot}$\fi} 
\def\ergscms{ergs~cm$^{-2}$~s$^{-1}$}  
  
\def\1755{4U~1755$-$33}
\def\AW{Angelini and White}
\def\xmm{{\it XMM-Newton}}
\def\chandra{{\it Chandra}}
\def\rxte{{\it RXTE~}}
\def\fas{\hbox{$.\!\!''$}}
\def\fam{\hbox{$.\!\!'$}}
\def\fad{\hbox{$.\!\!^\circ$}}

\shorttitle{Diffuse Emission in \1755}
\shortauthors{Park et al.}

\begin{document}

\title{A CHANDRA OBSERVATION OF THE DIFFUSE EMISSION CENTERED ON THE
LOW MASS X-RAY BINARY \1755}

\author{Shinae~Q.~Park\altaffilmark{1}, 
        Jon~M.~Miller\altaffilmark{1,2},
        Jeffrey~E.~McClintock\altaffilmark{1}, 
	Stephen~S.~Murray\altaffilmark{1}}

\altaffiltext{1}{Harvard--Smithsonian Center for Astrophysics, 
                 60 Garden Street, Cambridge, MA 02138; 
                 spark@cfa.harvard.edu, jmmiller@cfa.harvard.edu,
                 jmcclintock@cfa.harvard.edu, smurray@cfa.harvard.edu} 
\altaffiltext{2}{NSF Astronomy and Astrophysics Fellow} 

\begin{abstract}  

We present an analysis of a \chandra~observation of the field
surrounding the low-mass X-ray binary 4U 1755$-$33, which has been in
quiescence since 1996.  In 2003, Angelini \& White reported the
appearance of a narrow 7\arcmin long jetlike feature centered on the
position of 4U 1755$-$33 using the \xmm~telescope.  Though the source
and jet are not visibly apparent in our \chandra~ACIS-S image, there
is a significant excess (4--6$\sigma$) of counts in a region that
encloses the jet when compared to adjacent regions.  We examined a
knot of emission in the jet that was detected by \xmm~but not by
\chandra~and calculated that if the knot flux observed by
\xmm~was concentrated in a point source, \chandra~would have easily
detected it; we therefore conclude that this knot of emission is
diffuse.  In summary, we suggest that the jetlike feature found
previously in the \xmm~data is quite diffuse and likely a true jet,
and is not due to a chance alignment of discrete point sources or
point-like regions of emission associated with internal shocks.

\end{abstract}


\maketitle 


\section{Introduction} 

\1755 is a low-mass X-ray binary (LMXB) and black hole candidate
located in the direction of the Galactic center.  It was first
discovered by the {\it Uhuru} satellite's all-sky X-ray survey in 1970
\citep{gia74} with a flux of $\sim 100 \mu$Jy.  The X-ray source
remained bright and persistent until January 1996, when the {\it Rossi
X-ray Timing Explorer} (\rxte) failed to detect the source while in
its off state with a flux of $\le 1 \mu$Jy \citep{rob96}.  The source
has remained quiescent since then, to the time of writing.

The X-ray spectrum of \1755 indicates that the primary may be a black
hole.  The spectrum was observed to be ``ultrasoft'' ($kT \sim$ 2 keV)
when the source was bright \citep{jon77, whi84, whietal84}, and a hard
X-ray tail above $\sim 6-10$ keV was  observed by \citet{pan95}; both
spectral aspects are frequently observed for black hole candidates
\citep{tan95}.  \citet{seo95} also noted the appearance of an iron
emission line centered around 6.7 keV, as found in many black hole
candidates \citep{mil02, miller02, par04}.

The optical counterpart of \1755 was identified as a faint, blue star
with a featureless spectrum by \citet{mcc78}.  When the X-ray source
was active, the optical counterpart was a V$\sim$18--19 mag object.
Since turning off, the counterpart has dimmed to V $>$ 22
\citep{wac98}.  A regular, periodic dipping in the X-ray light curve
of the source was observed by \citet{whi84}, suggesting that the
source  is being observed at a high inclination angle.  This
periodicity also  allowed for the determination of the 4.4 hr orbital
period of the binary \citep{whietal84}, a result confirmed by the
photometric variations in the optical counterpart by \citet{mas85}.
This period, along with the period-mass relation \citep{fra02},
suggests a mass of $\sim$ 0.5 \msun~for the secondary star.  The
distance to this source is thought to be between 4--9 kpc
\citep{wac98}.

In 2003, \citeauthor{ang03} reported the \xmm~discovery of an
X-ray-emitting jetlike feature that is 7\arcmin in extent, narrow, and
centered at the position of \1755.  They suggest that it has existed
since the source entered quiescence.  The 3\fam5 half-length of the
jet corresponds to 4~pc for an assumed distance of 4~kpc.  Thus, as
\citeauthor{ang03} note, it is plausible that a relativistic jet could
have expanded to this length, given that the source was active for at
least 25 years.  They also report the existence of a hole developing
near the LMXB, which is consistent with the scenario that the source
has not been feeding the jet during the past eight years of
quiescence.  Although jets are not seen exclusively in black holes
binaries, they are relatively common in such systems
\citep[e.g.][]{mir03}.  Confirmation that this extended X-ray emission
feature is a true jet would provide further evidence that \1755 may
contain a black hole primary.  In this Letter, we present a
\chandra~observation of \1755 and report on the diffuse appearance of
the jetlike feature.

\section{Observations and Analysis} 

We observed \1755 with the {\it Chandra} X-ray Observatory on 2003
September 25 from 17:21:18 to 23:51:50 UT, with 21909 s of good time
exposure.  The incoming X-ray flux was read out by the Advanced CCD
Imaging Spectrometer (ACIS) spectroscopic array (ACIS-S) with the very
faint format in timed exposure mode.  At the time of the proposal, it
was not known that \1755 might have a jetlike feature extending from
it, and thus the source was placed at the standard aimpoint near the
boundary between nodes 0 and 1 on the ACIS-S3 CCD chip.  This aimpoint
is not at the center of chip, so unfortunately, the S3 CCD did not
capture the full length of the region of extended emission.  In order
to avoid measurement errors due to the very different responses of
the S3 and S2 chips, we chose to only look at ACIS-S3, 
cutting off about 15\% of the region in question (see Fig.\ 1).

\begin{center}
\psfig{file=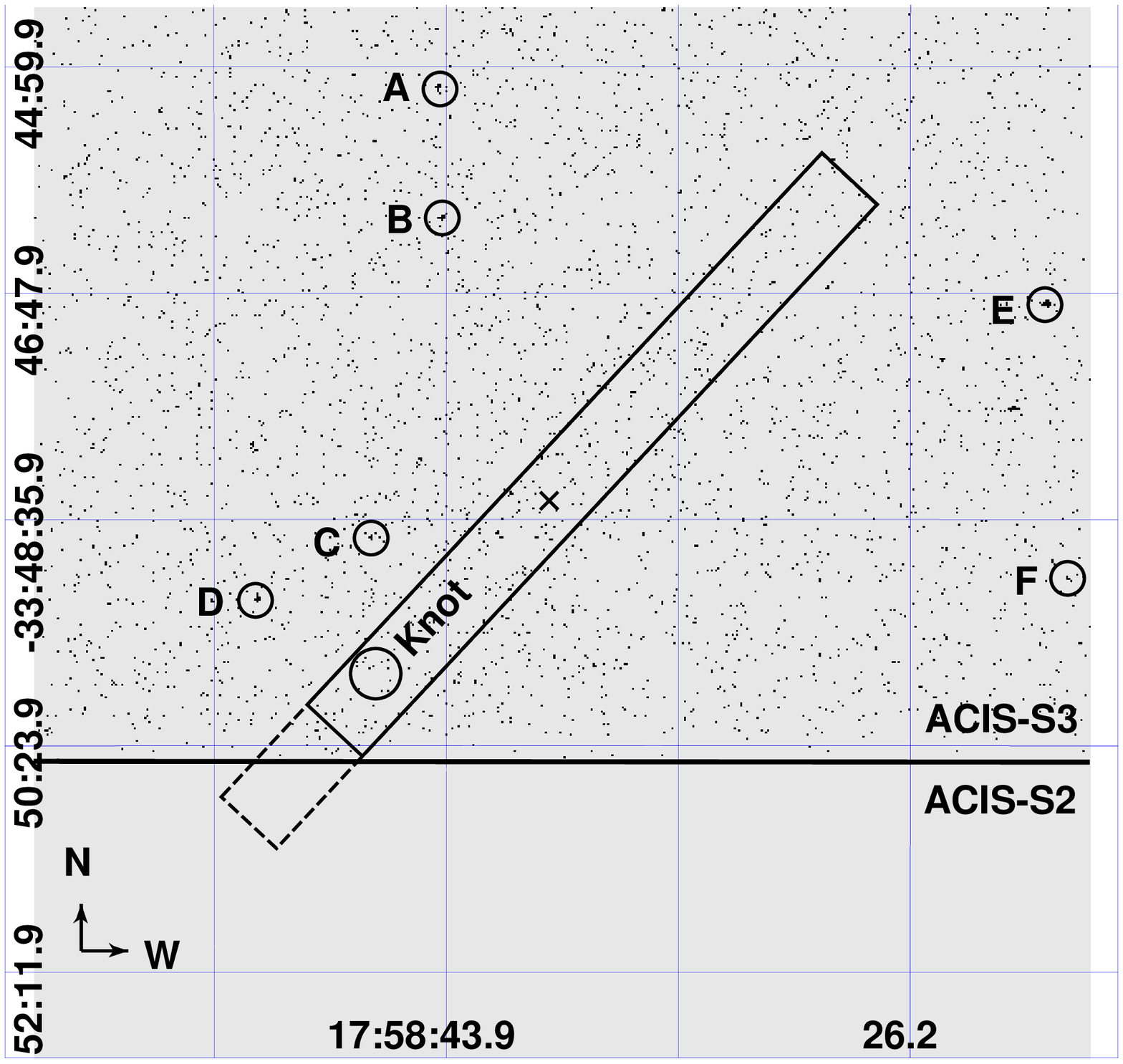, height=2.0in}
\figcaption{\chandra~ACIS-S3 chip image of the \1755 region in the
  0.3--7.0 keV band.  The full rectangular region of length 7\arcmin
  in the center of the image shows the extent of the jetlike emission
  observed by \citet{ang03} with \xmm; the central ``x'' denotes the
  position of \1755.  In the \chandra~analysis, we consider only the
  upper 6\arcmin of the emission (solid line) that was imaged by the
  ACIS-S3 chip.  The circle within the rectangular region encloses the
  location of the knot-like feature observed by \xmm.  Neither
  this feature nor the jet itself is apparent in a visual inspection
  of the ACIS-S3 image.  The smaller circles denoted by A-F are point
  sources detected by both \chandra~and \xmm.}
\end{center}

We compare the \chandra~data with the \xmm~data for the same
source, which were taken on 2001 March 8 in a 5 hour exposure using the two
EPIC-MOS arrays operated in full frame mode and the EPIC-PN array in
extended full frame mode \citep{ang03}.  All three cameras observed
the source with the ``medium'' optical blocking filter.  The good time
exposure in each camera was 18127 s, 18107 s, and 15359 s for the
MOS-1, MOS-2, and PN cameras respectively.

The \chandra~data were analyzed using standard processing tools from
CIAO version 3.0.2.  No significant flaring was seen in the data, and the
response files used were corrected for the gradual degradation in the
low-energy response of the ACIS detector.  We also reanalyzed the
\xmm~data obtained through the HEASARC public data archive using the
\xmm~Science Analysis System (SAS) version 5.4.1 to undertake a proper
comparison of the data between the two telescopes.  Because the PN
camera has less uniform coverage due to the gaps between the CCDs and
bad pixel columns, we chose to use only the data from the two MOS
cameras in our analysis.  The difference in sensitivity between these
two cameras is modest, and in our results we found that the MOS-2
camera yielded greater flux values by at most 15\%.  For our analysis,
we quote only the average of the two cameras.

The spectra of the data from both \chandra~and \xmm~were binned to
contain a minimum of 15 counts in each channel and analyzed in the
0.3--7.0 keV band using XSPEC version 11.2 \citep{arn00}. The region
of the jetlike feature was specified to be 7\arcmin in length by \AW,
and estimated by us to be 0\fam6 in width from the \xmm~images.
Because a feature 7\arcmin in length centered on the source does not
fit in the \chandra~ACIS-S3 chip, we took the region length to be
6\arcmin in our analysis of both the \chandra~and \xmm~data (see Fig.\
1) so that we could compare the fluxes using the exact same region.
For consistency, we always use the same source and background regions
in our analysis of both data sets.  All errors on fluxes reported in
this work are 95\% confidence errors, errors on counts are at 1$\sigma$,
and all other errors are at 90\% confidence unless otherwise noted.

\section{Results}

In its quiescent state, \1755 remained undetected by \chandra.  We
obtain an upper limit for the flux of the LMXB by finding an average
of 1.05 background counts in a 2\fas0 radius circle, the size of the
aperture that is expected to contain 90\% of the flux for an on-axis
ACIS point source at 1.49 keV.  A source with less than five counts
has a 3$\sigma$ probability of being detected according to Poisson
statistics; using a power-law model (with parameters equal to those we
used to analyze the jetlike emission, described in detail below) with
WebPIMMS\footnote{http://heasarc.gsfc.nasa.gov/Tools/w3pimms.html},
this corresponds to a flux upper limit of $2.98 \times 10^{-15}$
\ergscms.

\subsection{Emission from the Jetlike Region}

In 2003, \AW~reported the detection of a 7\arcmin jetlike feature of
extended emission, centered on the position of \1755 and at a position
angle of $\sim 137\deg$.  \1755 is located off the galactic plane at
RA 17$^h$58$^m$40$.\!\!^s$0 and DEC -33\deg48$^\prime$27\fas0 ($l =
357\fad21$, $b = -4\fad87$), so the emission is not likely to be due
to a feature in the plane.  The emission is also found to be
symmetric, as expected from a binary source observed at high
inclination.

The jetlike feature is very prominent in the \xmm~images of Angelini
and White (see Fig.\ 1 and Fig.\ 2 of their paper).  In the
high-resolution \chandra~image, however, the jet is not visually
apparent.  The  \chandra~image provides evidence that the emission may
be diffuse: a rectangular area of 0\fam6 $\times$ 6\arcmin
enclosing the expected jetlike feature contains many
more counts than regions of identical area next to the jet region, as
shown in Figure 2.  The central region was found to have 1327$\pm$36
counts, which is 4.1 sigma above the region directly to the northeast
(with 1122$\pm$34 counts) and 5.8 sigma above the region directly to
the southwest (with 1044$\pm$32 counts).

\begin{center}
\psfig{file=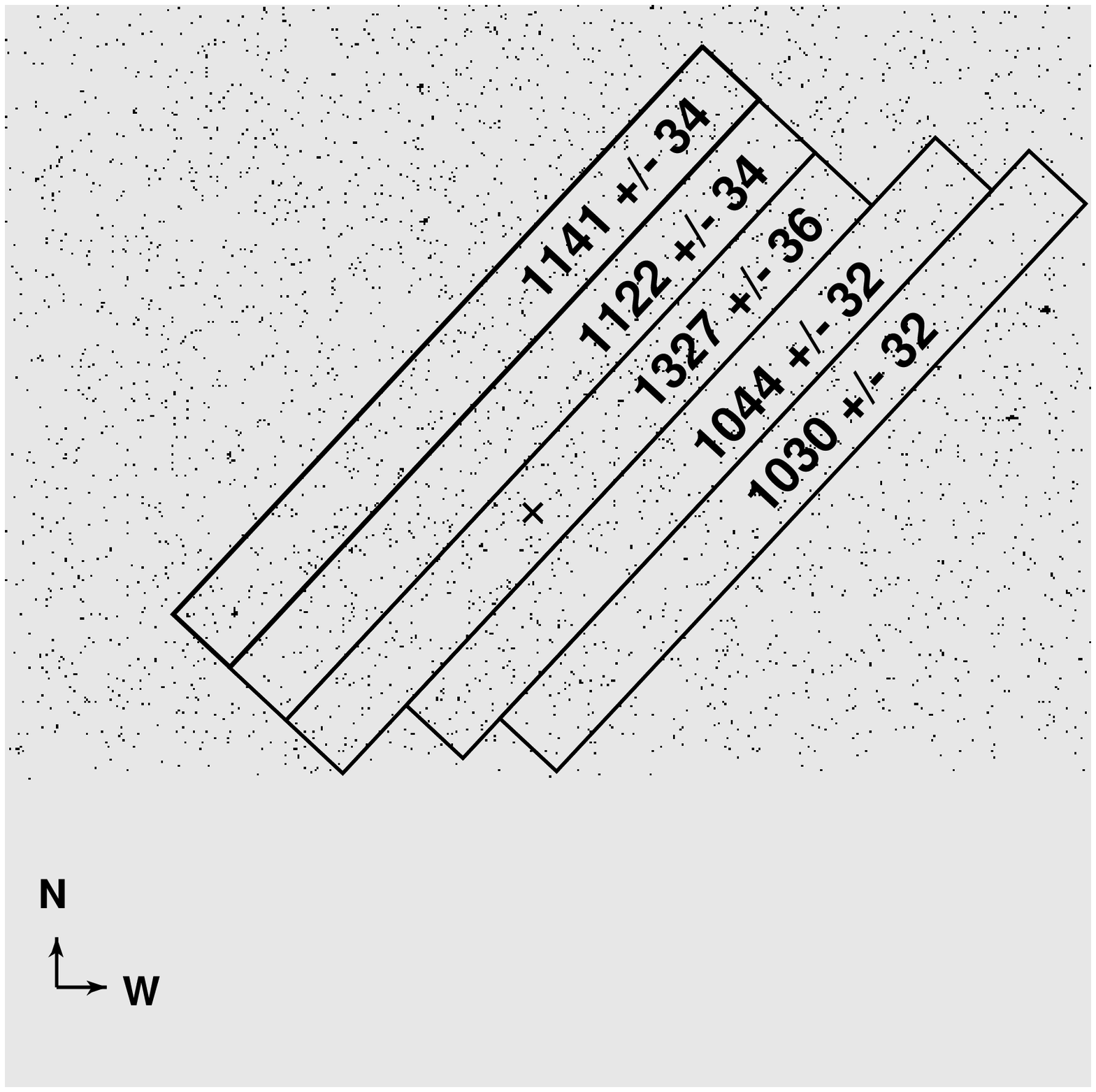, height=2.0in} 
\figcaption{\chandra~region counts.  The central rectangle, which is
  identical to the solid-line rectangle shown in Figure~1, encloses
  most of the jetlike feature observed with \xmm.  The four flanking
  rectangles enclose background comparison regions.  The total number
  of counts detected after point source subtraction in each of the
  0\fam6 $\times$ 6\arcmin regions is indicated.  As before, the
  ``x'' in the central rectangle marks the location of \1755.
  The same rectangular regions were used in analyzing the \xmm~data.}
\end{center}

The jet region has $244.4 \pm 43.6$ net counts using the two outer
rectangles as the background region.  (The lower 0\fam6 of the
leftmost rectangle was excluded due to contamination by  point source
D -- see Fig.\ 1).  This background region is used for all  analysis
in this paper.  (See Table 1 for a description of the counts recorded
in various regions.)  The local background appears to be relatively
independent of location in the \xmm~data, whereas there is a
significant variation in the \chandra~background on the two sides of
the central rectangle that contains the source. Spatial variations of
this magnitude are frequently observed in ACIS
images\footnote{http://cxc.harvard.edu/contrib/maxim/bg/index.html}.
The standard ACIS-S3 background dataset from  the calibration database
(CALDB) showed the same ratio of variation between the sum of the two
leftmost rectangles and the sum of the two rightmost rectangles,
suggesting that the variation we see is not specific to our image.

We adopted the simple power-law and thermal bremsstrahlung models used
by \AW.  For both models, we froze the hydrogen column density at
$N_{\rm H}$ = $3 \times 10^{21}$ cm$^{-2}$, which is the value
predicted from H I maps in the direction and at the distance of \1755
\citep{dic90, fru94}, and is similar to the $N_{\rm H}$ values adopted
by others \citep[e.g.][]{seo95}.  From our best fit power-law model,
we obtained photon indices of $1.2 \pm 0.4$ for the \chandra~data, and
$1.6 \pm 0.2$ for the \xmm~data.  These are also consistent with the
photon index of $1.8 \pm 0.3$ found by \citet{ang03} using the same
value of $N_{\rm H}$.  In the various models they adopt, \AW~find
photon indices ranging from 1.5 to 1.9 for their jet region, which
they note are similar to values observed from other galactic and
extragalactic jets associated with black holes.  In order to better
constrain our model, we chose to freeze the power-law photon index at
1.5 in our determination of the flux.  The temperature was also very
poorly constrained in the bremsstrahlung model.  Because we could not
get a precise value for this parameter, and because \AW~report a lower
limit of 4 keV in their analysis, we chose to constrain our model by
freezing the temperature at 4 keV.  Fixing the temperature at 4 keV or
at 10 keV returned minimal differences in the fluxes found.

Our fits for the normalization parameter generated a reduced $\chi^2$
of 1.03 and 0.30 for both models in \chandra~and \xmm, respectively.
The 0\fam6 $\times$ 6\arcmin jet region in the \xmm~image yielded an
average unabsorbed flux of $0.96 \pm 0.34 \times 10^{-13}$
\ergscms~for the MOS-1 and MOS-2 cameras and $1.32 \pm 0.45 \times
10^{-13}$ \ergscms~in \chandra~in the 0.3--7.0 keV band with the
power-law model.  (See Table 2 for a summary of these results and the
results from the bremsstrahlung model.)  These fluxes correspond to a
luminosity of $1.83 \pm 0.65 \times 10^{32}~d^2_{4 kpc}$ ergs~s$^{-1}$
from the \xmm~data and $2.51 \pm 0.86 \times 10^{32}~d^2_{4 kpc}$
ergs~s$^{-1}$ from the \chandra~data.

\subsection{Emission in the Knot Region}

Two bright knotlike regions were found in the \xmm~image in the
southeastern end of the emission area.  They appear to be broader than
the point-spread function (PSF) of \xmm~of 9\arcsec in radius at 1.5
keV for a region encircling half of the total energy.  One of these
regions was also imaged by {\it Chandra} (see Fig.\ 1).
We analyzed this knot, at RA 17$^h$58$^m$46$.\!\!^s$64 and DEC
-33\deg49$^\prime$49\fas48, using the same models described above.  We
selected a 24\arcsec diameter extraction aperture by comparing the net
counts for a series of aperture sizes and noting where the flux fell
off most significantly.  The background region used was identical to
the one described in our jet analysis.  

The knot region had a greater concentration of net counts compared to
the net counts in the region of jetlike emission in the \xmm~data, but
the knot emission was not detected at a significant level in the
\chandra~data.  \xmm~recorded $34.4 \pm 7.2$ net counts in the knot,
which, when scaled for area, is greater than the average net counts in
the jet region significant to 3.5$\sigma$.  \chandra~recorded $12.2
\pm 7.1$ net counts, which is not statistically different than the net
counts recorded in the jet region.  The unabsorbed 0.3--7.0 keV flux
with the power-law model yielded an upper limit of $4.63 \times
10^{-14}$ \ergscms~from the averaged MOS cameras and $1.32 \times
10^{-14}$ \ergscms~from the \chandra~data.  See Table 1 for a
description of the counts obtained in the knot region, and Table 2 for
a summary of the fluxes found obtained from both the power-law and the
bremsstrahlung models for the knot region.

\section{Discussion}

The lack of visibility of the knot in \chandra~suggests that the
object may be diffuse.  We verified this hypothesis by comparing the
ratio of the fluxes between the knot and the jet in \xmm~and \chandra.
Assuming that $N_{\rm H}$ and $\Gamma$ are the same for both the knot
and the jet, and that the flux of the knot has not changed
significantly in the 2.5 years between the \xmm~and
\chandra~observations, the ratio between the net knot and jet flux in
\xmm~gives a predicted 30.3 $\pm$ 9.0 net counts for the knot in
\chandra~for the 0.3--7.0 keV band.  This method of scaling gives a
count value for the knot that is different from the actual 12 $\pm$
7.1 net counts of observed in the \chandra~image by 1.58 $\sigma$.
Choosing harder energy bands makes the knot more significant: in the
4.0--7.0 keV band, the knot net counts is higher than the jet net
counts by 3.0 $\pm$ 0.5; for the 0.3--7.0 keV band, the knot is only
1.4 $\pm$ 0.4 times higher.  The surface brightness sensitivity of
\chandra~may be too low to statistically detect the a diffuse knot in
the 0.3--7.0 keV band.  If all of the counts in the knot had been
concentrated at a point source, the knot would be significant and
immediately apparent in the \chandra~image, since the  background in a
typical 2\fas0 radius extraction aperture is only 1.05 counts in the
0.3--7.0 keV band (see Table 1).   Thus, we suggest that the observed
knot is most likely a real, but diffuse, feature.

Likewise, the jetlike emission is also likely a real feature.  The
agreement between the \chandra~and \xmm~fluxes for the jet region
suggest that the emission visibly seen in the \xmm~data must still be
present in the \chandra~data.  Though the \chandra~data had slightly
longer good time exposure ($\sim$ 22000 s, compared to $\sim$ 18100 s
for the MOS cameras aboard \xmm), there are other factors that have
contributed to the lack of visibility of the jet in \chandra,
including \xmm's greater effective area: at 1.5 keV, the effective
area of the MOS cameras aboard \xmm~is 1400 cm$^{2}$ compared to
\chandra's 700 cm$^{2}$.  The background rates are also twice as high
with \chandra, making significant detections of diffuse emission more
difficult.  Thus, if
the emission were truly diffuse, it would have been more easily
detected with the capabilities of the \xmm~instruments.

The emission is not likely to be due to an alignment of point sources,
as multiple point sources were not found in the emission region by
either \chandra~or \xmm.  Six point sources located outside the jet
region, which were detected by both \chandra~and \xmm, are shown in
Figure 1 with their positions and net counts listed in Table 3. Of
course, a large enough assemblage of very faint point sources would
have escaped detection as individual sources (e.g., 100 or more source
with fluxes of $\sim 10^{-15}$ \ergscms~or less), although this
scenario is an unlikely one.

It is important to note that since one of the bright knotlike regions
that was detected by \xmm~was off the S3 chip of {\it Chandra}, a
simple scaling of the fluxes found in our analysis of the 0\fam6
$\times$ 6\arcmin region to the estimated true size of the jet of
7\arcmin may yield a slight underestimate of the actual flux in the
region.  However, since we cannot analyze the \chandra~image with a
7\arcmin jet, we limited our \xmm~analysis to 6\arcmin regions for
more accurate and consistent comparisons between the two telescopes.


\section{Conclusions}

While the \chandra~image showed many point sources, emission from LMXB
\1755 in quiescence was not detected.  We present the results of a
comparative study using \chandra~and \xmm~imaging data of a jetlike
emission feature that is apparently emanating from \1755.  Although
the jetlike feature is not directly visible in the \chandra~image,
though highly visible in the \xmm~images, \chandra~does detect
significant emission (4--6$\sigma$) from the jet region.  Furthermore,
we find that the fluxes from the same 0\fam6 $\times$ 6\arcmin jet
region as observed by \chandra~and \xmm~agree within errors.  The flux
of a knotlike region embedded in the jet was detected by \xmm, but not
by \chandra~because of \chandra's smaller area and higher background.
We show that if this emission were due to a point source, then it
would have been easily detected by \chandra.  We therefore conclude
that the emission from the knot is diffuse.  Our analysis shows that
the jetlike emission surrounding \1755 is likely a true and diffuse
jet, which appears to be devoid of point-like regions of emission
associated with internal shocks.  Because jets evolve very rapidly in
our galaxy, compared with the slowly evolving larger scale jets of
quasars and AGN, studying the jets from \1755 and other galactic
sources may provide key information in understanding jet formation and
propagation.

\acknowledgments We thank W. Forman and M. Markevitch for discussions
on the spatial structure of the ACIS background.
J.~M.~M.~acknowledges support from the NSF through its Astronomy and
Astrophysics Postdoctoral Fellowship program.  This work has made use
of the information and tools available at the HEASARC Web site,
operated by GSFC for NASA.

\begin{table}
\begin{center}
\begin{tabular}{cccccccc}
\multicolumn{8}{c}{Table 1: {\textbf{Region Counts}}}\\
\hline  
\hline 
~~ & \multicolumn{2}{c}{JET} && \multicolumn{2}{c}{KNOT} && BACKGROUND\\ 
\cline{2-3} \cline{5-6} \cline{8-8}
      & Total Counts & Net Counts && Total Counts & Net Counts  && Total 
Counts\\
\hline
\xmm      & 782.0$\pm$28.0  & 277.8$\pm$32.4 && 52.0$\pm$7.2 & 34.4$\pm$7.2 && 958.0$\pm$31.0  \\
\chandra  & 1327.0$\pm$36.4 & 244.4$\pm$43.6 && 50.0$\pm$7.1 & 12.2$\pm$7.1 && 2057.0$\pm$45.4\\
\cline{2-3} \cline{5-6} \cline{8-8}
Region Size & \multicolumn{2}{c}{$0\fam6 \times 6^{\prime}$} && \multicolumn{2}{c}{$\pi \times 0\fas2^{2}$} && $(0\fam6 \times 6^{\prime}) + (0\fam6 \times 5.4^{\prime}$)\\
\tableline
\end{tabular}
\end{center}
\tablecomments{Counts for the jet, knot, and background regions of
both telescopes in the 0.3--7 keV band.  The \xmm~information is from
a summed image of MOS 1 and MOS 2.  The counts in the jet and knot
regions were found using the central rectangular region and circular
region in Figure 1, respectively.  The outer two regions in Figure 2
were adopted as the background regions for both the knot and the jet,
with the leftmost rectangle truncated to the upper 5\fam4 to avoid
contamination from counts from a point source.}
\end{table}

\begin{table}
\begin{center}
\begin{tabular}{lcc}
\multicolumn{3}{c}{Table 2: {\textbf {Flux Limits For The Jet And Knot Regions}}}\\
\hline \hline
~~ & JET REGION & KNOT REGION\\ 
\hline
\multicolumn{3}{c}{Power-law Model}\\
\hline
\chandra  & $~~1.32 \pm 0.45 \times 10^{-13}~~$ & $~1.32 \times 10^{-14}~$  \\
\xmm      & $~~0.96 \pm 0.34 \times 10^{-13}~~$ & $~4.63 \times 10^{-14}~$  \\
\hline					       
\multicolumn{3}{c}{Bremsstrahlung Model} \\
\hline
\chandra  & $~~1.07 \pm 0.37 \times 10^{-13}~~$ & $~1.07 \times 10^{-14}~$  \\
\xmm      & $~~0.86 \pm 0.31 \times 10^{-13}~~$ & $~4.01 \times 10^{-14}~$  \\
\hline
\end{tabular}
\end{center}
\tablecomments{All fluxes are listed with  95\% confidence errors in units of
  \ergscms in the 0.3--7.0 keV band.  Because the errors in the knot
  flux are large, we list only the upper limits for the knot region.}
\end{table}

\begin{table}
\begin{center}
\begin{tabular}{cccc}
\multicolumn{4}{c}{Table 3: {\textbf{Point Source Positions}}}\\
\hline  
\hline 
ID & RA & DEC & Net Counts \\
\hline
~~~A~~~  & ~~17 58 44.24~~  & -33 45 09.52 & 77 \\
~~~B~~~  & ~~17 58 44.06~~  & -33 46 11.83 & 54 \\
~~~C~~~  & ~~17 58 46.81~~  & -33 48 44.74 & 43 \\
~~~D~~~  & ~~17 58 51.17~~  & -33 49 13.58 & 89 \\
~~~E~~~  & ~~17 58 20.99~~  & -33 46 53.50 & 97 \\
~~~F~~~  & ~~17 58 20.15~~  & -33 49 04.03 & 15 \\
\tableline
\end{tabular}
\end{center}
\tablecomments{Positions of the six point sources detected by both \chandra~and \xmm~
(epoch 2000).  See Figure 1 for an image.  The net counts were taken
from the \chandra~observation.}
\end{table}


\begin{thebibliography}{}
\expandafter\ifx\csname natexlab\endcsname\relax\def\natexlab#1{#1}\fi
\bibitem[Angelini \& White (2003)]{ang03} Angelini, L., \& White, N.~E.\ 2003, \apj, 586, L71
\bibitem[Arnaud \& Dorman (2000)]{arn00} Arnaud, K.~A., \& Dorman, B.\ 2000, XSPEC is available via the HEASARC on-line service, provided by NASA/GSFC   
\bibitem[Dickey \& Lockman (1990)]{dic90} Dickey, J.~M. \& Lockman, F.~J.\ 1990, \araa, 28, 215
\bibitem[Frank, King, \& Raine (2002)]{fra02} Frank, J., King, A., \& Raine, D.~J.\ 2002, Accretion Power in Astrophysics (3d ed.; Cambridge: Cambridge University Press), 398
\bibitem[Fruscione et al.\ (1994)]{fru94} Fruscione, A., Hawkins, I., Jelinsky, P., \& Wiercigroch, A.\ 1994 \apjs, 94, 127
\bibitem[Giacconi et al.\ (1974)]{gia74} Giacconi, R., Murray, S., Gursky, H., Kellogg, E., Schreier, E., Matilsky, T., Koch, D., \& Tananbaum, H.\ 1974, \apjs, 27, 37
\bibitem[Jones (1977)]{jon77} Jones, C.\ 1977, \apj, 214, 856
\bibitem[Mason, Parmar, \& White (1985)]{mas85} Mason, K.~O., Parmar, A.~N., \& White, N.~E.\ 1985, \mnras, 216, 1033
\bibitem[McClintock et al.\ (1978)]{mcc78} McClintock, J., Canizares, C., Hiltner, W.~A., Petro, L., \& Griffiths, R.\ 1978, IAU Circ., 3251, 1
\bibitem[Miller et al.\ (2002a)]{mil02} Miller, J.~M., et al.\ 2002a, \apj, 570, L69
\bibitem[Miller et al.\ (2002b)]{miller02} Miller, J.~M., et al.\ 2002b, \apj, 578, 348
\bibitem[Mirabel (2003)]{mir03} Mirabel, I.~F.\ 2003, New Astronomy Review, 47, 471
\bibitem[Pan et al.\ (1995)]{pan95} Pan, H.~C., Skinner, G.~K., Sunyaev, R.~A., \& Borozdin, K.~N.\ 1995, \mnras, 274, L15
\bibitem[Park et al.\ (2004)]{par04} Park, S.~Q., et al.\ 2004, \apj, 610, 378
\bibitem[Roberts et al.\ (1996)]{rob96} Roberts, M.~S.~E., Michelson, P.~F., Cominsky, L.~R.,Marshall, F.~E., Corbet, R.~H.~D., \& Smith, E.~A.\ 1996, IAU Circ., 6302, 2
\bibitem[Seon et al.\ (1995)]{seo95} Seon, K., Min, K., Yoshida, K., Makino, F., van der Klis, M., van Paradijs, J., Lewin, W.~H.~G.\ 1995, \apj, 454, 463
\bibitem[Tanaka \& Lewin (1995)]{tan95} Tanaka, Y., \& Lewin, W.~H.~G.\ 1995, X-Ray Binaries, ed. W.~H.~G.~Lewin, J.~van Paradijs, \& E.~P.~J.~van den Heuvel (Cambridge: Cambridge Univ. Press), 126
\bibitem[Wachter \& Smale (1998)]{wac98} Wachter, S., \& Smale, A.~P.\ 1998, \apj, 496, L21
\bibitem[White \& Marshall (1984a)]{whi84} White, N.~E., \& Marshall, F.~E.\ 1984a, \apj, 281, 354
\bibitem[White et al.\ (1984b)]{whietal84} White, N.~E., Parmar, A.~N., Sztajno, M., Zimmermann, H.~U., Mason, K.~O., \& Kahn, S.~M.\ 1984b, \apj, 283, L9
\bibitem[Wise et al.\ (2003)]{wis03} Wise, M.~W., Davis, J.~E., Huenemoerder, D.~P., Houck, 	J.~C., \& Dewey, D.\ 2003, MARX 4.0 Technical Manual
\end{thebibliography}
\end{document}